\documentclass[aps,prb,floatfix,superscriptaddress,twocolumn]{revtex4-2}
\usepackage[utf8]{inputenc}
\usepackage[canadian]{babel}
\usepackage{graphicx}
\usepackage{amsmath}
\usepackage{amsthm}
\usepackage{amsfonts}
\usepackage{amssymb}
\usepackage{upgreek}
\usepackage{bm}
\usepackage{soul}
\usepackage{textcomp}
\usepackage{xcolor}
\usepackage{booktabs}
\usepackage{array}
\usepackage[pdfusetitle]{hyperref}
\usepackage[caption=false]{subfig}
\usepackage[capitalize]{cleveref} 
\usepackage[version=3]{mhchem}
\usepackage{enumitem}
\usepackage{ifthen}
\usepackage[ISO]{diffcoeff}[=v4]
\usepackage{siunitx} 

\newcommand{\phantomsubfloat}[1]{
  {%
    \captionsetup[subfigure]{labelformat=empty}
    \subfloat[][]{#1}
  }%
}
\newcolumntype{L}{>{\(}l<{\)}} 
\hypersetup{
  hypertexnames=false,
  colorlinks=true,
  linkcolor=blue,
  anchorcolor=blue,
  citecolor=blue,
  filecolor=blue,
  urlcolor=blue,
  pdfauthor={Igor Benek-Lins; Saurabh Maiti},
}

\newcommand{\beq}{\begin{equation}}
\newcommand{\eeq}{\end{equation}}
\newcommand{\bea}{\begin{eqnarray}}
\newcommand{\eea}{\end{eqnarray}}
\long\def\bal#1\eal{\begin{align}#1\end{align}}
\newcommand{\bse}{\begin{subequations}}
\newcommand{\ese}{\end{subequations}}
\newcommand{\nn}{\nonumber}
\newcommand{\bwt}{\begin{widetext}}
\newcommand{\ewt}{\end{widetext}}

\DeclareDocumentCommand\differential{ o g d() }{%
  \IfNoValueTF{#2}{
    \IfNoValueTF{#3}
    {\dl\IfNoValueTF{#1}{}{^{#1}}}
    {\mathinner{\dl\IfNoValueTF{#1}{}{^{#1}}\argopen(#3\argclose)}}
  }
  {\mathinner{\dl\IfNoValueTF{#1}{}{^{#1}}#2} \IfNoValueTF{#3}{}{(#3)}}
}
\DeclareMathOperator{\diag}{diag}
\DeclareMathOperator{\sgn}{sgn}

\DeclareMathOperator{\IM}{Im}
\newcommand{\im}{\ensuremath{{\mkern1mu\mathrm{i}\mkern1mu}}}
\newcommand{\variation}{\updelta}

\newcommand{\spp}[1][]{\ensuremath{s_{++}}}
\newcommand{\spm}[1][]{\ensuremath{s_{+-}}}
\newcommand{\tD}[1][]{%
  \ifthenelse{\equal{#1}{}}{\ensuremath{2\Delta}}{\ensuremath{2\Delta_{#1}}}}
\newcommand{\tDabs}[1][]{%
  \ifthenelse{\equal{#1}{}}{\ensuremath{2|\Delta|}}{\ensuremath{2|\Delta_{#1}|}}}
\newcommand{\Dabs}[1][]{%
  \ifthenelse{\equal{#1}{}}{\ensuremath{|\Delta|}}{\ensuremath{|\Delta_{#1}|}}}
\DeclareDocumentCommand\order{ l m }{\fbraces#1{\lparen}{\rparen}{\mathcal{O}}{#2}}

\DeclareDocumentCommand\dd{}{\differential}
\newcommand{\group}[1]{\ce{#1}}
\newcommand{\Bog}{\group{B_{1g}}}

\newcommand{\Aog}{\group{A_{1g}}}

\newcommand{\BaKFeAsX}{\ce{Ba_{1-$x$}K_{$x$}Fe2As2}}

\newcommand{\LiFeOHFeSeX}{\ce{Li_{1-$x$}Fe_{$x$}OHFeSe}}
\newcommand{\MgBt}{\ce{MgB2}}

\newcommand{\swave}{\(s\)-wave}

\begin{document}
\title{Many-body physics-induced selection rules: application to Raman spectroscopy}
\author{Igor Benek-Lins}
\author{Saurabh Maiti}
\affiliation{Department of Physics, Concordia University, Montréal, QC H4B 1R6, Canada}
\date{\today}

\begin{abstract}
Spectroscopic measurements in quantum systems are subject to selection rules, usually based on space-time symmetries, that allow or disallow transitions between states. In many-body systems, in addition to the single-particle states, there emerge new ones due to collective excitations of the system. Here we demonstrate the existence of a ``fragile'' selection rule that emerges as a manifestation of many-body effects and outlines the conditions for collective excitations to couple to a given spectroscopic probe beyond the usual symmetry considerations. As an example, we apply the rule to Raman spectroscopy of multiband superconductors and settle some unresolved features in experiments.
\end{abstract}

\maketitle

\paragraph*{Introduction.} Single and two-particle (collective) excitations in quantum matter are important characteristics that provide a glimpse into how the system interacts with its environment or an external probe. A thorough understanding of these excitations is essential to interpret experimental data and infer valuable properties of quantum materials. Whilst the single-particle properties are readily derived from the knowledge of the electronic structure, the collective excitations require more detailed modelling involving an understanding of many-body correlations.
However, when discussing the coupling of these modes to a given probe, the many-body effects are often ignored under the assumption that they have already been incorporated in figuring out the collective modes. Here, we explicitly point out important modifications to selection rules from many-body effects.

While every probe would need its own consideration, in this work we shall study non-resonant electronic Raman spectroscopy (eRS), which is a powerful tool to probe collective modes in many irreducible representations (irreps) of a lattice simultaneously \cite{shastry_shraiman:1990:TheoryRamanScatteringMottHubbard,devereaux_hackl:2007:InelasticLightScatteringCorrelated}. Its application to superconductivity (SC), for instance, has enjoyed a long history from studying the coupling of SC collective excitations to phonons \cite{balseiro_falicov:1980:PhononRamanScatteringSuperconductors,littlewood_varma:1982:AmplitudeCollectiveModesSuperconductors,klein:1982:TheoryRamanScatteringChargedensitywave}, tracking the order parameter, $\Delta$, evolution and detecting higher angular momentum pairing symmetry (in cuprates) \cite{dierker_etal:1983:ElectronicRamanScatteringSuperconductingGap,hackl_etal:1983:GapModeSuperconductingGap,devereaux_etal:1994:ElectronicRamanScatteringHigh}, detecting multiband collective modes (\MgBt{}'s Leggett mode \cite{leggett:1966:NumberPhaseFluctuationsTwoBandSuperconductors}) \cite{blumberg_etal:2007:ObservationLeggettCollectiveMode,chubukov_etal:2009:TheoryRamanResponseSuperconductor,klein:2010:TheoryRamanScatteringLeggett,cea_benfatto:2016:SignatureLeggettModeGroupA}, to affirming the role of spin fluctuations in the pairing mechanism in Fe-based SCs \cite{bohm_etal:2014:BalancingActEvidenceStrong,maiti_etal:2016:ProbingPairingInteractionMultiple,bohm_etal:2018:MicroscopicOriginCooperPairing} (by tracking Bardasis--Schrieffer (BaSh) modes \cite{bardasis_schrieffer:1961:ExcitonsPlasmonsSuperconductors}), to name a few.

Earlier theoretical efforts to model the eRS \cite{devereaux_etal:1996:MultibandElectronicRamanScattering,klein:2010:TheoryRamanScatteringLeggett,cea_benfatto:2016:SignatureLeggettModeGroupA} did not aptly account for the multiband nature of systems (except for \MgBt{}). Thus, some very prominent features in experiments, such as comparable spectral weights of both the coherent collective mode and the incoherent collective excitations near the $2|\Delta|$ region \cite{kretzschmar_etal:2013:RamanScatteringDetectionNearlyDegenerate,bohm_etal:2014:BalancingActEvidenceStrong,bohm_etal:2018:MicroscopicOriginCooperPairing,wu_etal:2017:SuperconductivityElectronicFluctuationsCeBa,he_etal:2020:RamanStudyCooperPairing} remains unresolved to date. These theoretical studies suggested that the coherent mode, when present, should ``steal'' all the spectral weight \cite{scalapino_devereaux:2009:CollectiveWaveExcitonModes,khodas_etal:2014:CollectiveModesMultibandSuperconductors}. To address this, we calculate the eRS spectrum for a general multiband system. We discover the existence of a new selection rule induced by the many-body correlations that is able to explain the distribution of spectral weights between the collective mode and the $2|\Delta|$ region as seen in experiments on \LiFeOHFeSeX{} and \BaKFeAsX{} in all irreps. We are also able to show, for the first time, that the eRS response is actually sensitive to the sign change of the order parameter and even the many-body interaction matrix elements.

\textit{Internal degrees of freedom and the probe.} When we encounter a phase transition, new quantum degrees of freedom emerge that correspond to fluctuations of the established order parameter. Many-body coherence brought to these fluctuations by some effective attractive interaction leads to formation of collective modes that could couple to a probe that is sensitive to these fluctuating components. For example, in a SC there are four types of fluctuations of interest: amplitude (call it sector~1), phase (sector~2), density (sector~3) and velocity (sector~0). The quantum make-up of these sectors can be captured by representing them using the $2\times2$ identity and Pauli matrices ${\sigma_0}$ and $\{\sigma_i\}$ for $i\in\{1,2,3\}$. If we consider probes that involve the $\sigma_3$ (density) sector, then they can probe the system's density sector and all the fluctuations coupled to it.
For optical conductivity, it is the sector 0 that is relevant but sector 3 also becomes relevant in the presence of supercurrent \cite{nakamura_etal:2019:InfraredActivationHiggsMode}.
If we represent with $\gamma^{R,\rm probe}_i$ some physical property of the system that couples directly to the probe through the sector $i$ in the $R^{\rm th}$ irrep (we call this the probe vertex), then the additional selection rules we find dictate which fluctuations couple to which combinations of $\gamma^{R,\rm probe}_i$.

\paragraph*{The many-body selection rule.} In this work, we will devise a rule for the coupling to the $\sigma_3$ sector which would be applicable to non-resonant eRS in any irrep \cite{devereaux_hackl:2007:InelasticLightScatteringCorrelated,lazarevic_hackl:2020:FluctuationsPairingFebasedSuperconductors}, THz spectroscopy \cite{fiore_etal:2023:ManipulatingPlasmaExcitationsTerahertz} and also to supercurrent-assisted optical absorption \cite{moor_etal:2017:AmplitudeHiggsModeAdmittance,nakamura_etal:2019:InfraredActivationHiggsMode}.
In a SC, it is the phase fluctuations $\variation\varphi$ (sector $\sigma_2$) of the order parameter \(\Delta\) that couples to sector $\sigma_3$ \cite{maiti_etal:2017:ConservationLawsVertexCorrections}. For a multiband system with $\Delta_{b_i}=|\Delta_{b_i}|e^{\im\varphi_{b_i}}$, where $i\in\{1,2,\ldots\}$ runs over the bands, we find that a collective mode in irrep $R$ with fluctuation form factor $\variation\varphi^R_{b_1}\pm s\variation\varphi^R_{b_2}$ couples to the probe via $\gamma^R_{3,b_1}\pm s\gamma^R_{3,b_2}$, respectively, where $s=-\sgn(V_{R,b_1b_2}^{\rm pp}\Delta_{b_1}\Delta_{b_2})$ and $V_{R,b_1b_2}^{\rm pp}$ is the interaction matrix element between bands $b_1$ and $b_2$ in the $R^{\rm th}$ irrep in the particle-particle interaction channel (pp) in which the instability (here SC) took place. Moreover, the mode corresponding to $\variation\varphi^R_{b_1}+s\variation\varphi^R_{b_2}$ would be the lower energy mode and that to $\variation\varphi^R_{b_1}-s\variation\varphi^R_{b_2}$ would be the higher energy mode. In what follows, we will fix our probe to be eRS and explicitly demonstrate the above points for \Aog{} and \Bog{} spectra in two-band systems. In the \Bog{} case, the rule becomes approximate and hence we use the adjective ``fragile'' to describe it. We even use the rule to address eRS from disconnected multi-Fermi-pocket models where multi-orbital physics becomes relevant. Its extension to 3-band systems and systems with broken time-reversal symmetry is provided in a concomitant work \cite{sarkar_maiti:2023:ElectronicRamanResponseSuperconductor}.

\paragraph*{Application to eRS.} Consider a system with $n$ bands with energy dispersion $\epsilon_{\vec k}^{b_i}$ and Raman vertex $\gamma^R_{b_i}$ projected onto the irrep $R$ (this plays the role of $\gamma^{R,\rm probe}_{n,b_i}$). Let us model the Cooper (pp) channel interaction as $\hat V^{\rm pp}=\hat V^{\rm pp}_{\Aog}f^{\Aog}_{\theta_{\vec k}}+\hat V^{\rm pp}_{\Bog}f^{\Bog}_{\theta_{\vec k}}$, where the form factors are $f^{\Aog}_{\theta_{\vec k}}=1$ and $f^{\Bog}_{\theta_{\vec k}}=\sqrt2\cos2\theta_{\vec k}$. $\theta_{\vec k}$ is the angle along the Fermi surface measured relative to the $\Gamma \group{X}$ axis and `$\hat~$' represents a matrix in the band space. We assume the leading channel to be \Aog{} so that the self-consistency equations [see the supplementary material (SM)] lead us to \((\Delta_{b_i}^{\Aog}, \Delta_{b_i}^{\Bog})=(\Delta_{b_i}, 0)\). The central object to study the collective modes is the correlation function between various degrees of freedom (sectors) $m,n\in\{0,1,2,3\}$:
\begin{align}\label{Eq:Pi33Pi32}
  \Pi^{b_i}_{mn}(\im\Omega_m)=\int_K{\rm Tr}[\hat\sigma_m\hat G^{b_i}_K\hat\sigma_n\hat G^{b_i}_{K+Q}],
\end{align}
where $K\equiv (\im\omega_n,\vec k)$, $Q\equiv (\im\Omega_m,0)$, $\hat\sigma_m$ are the Pauli matrices, acting on the particle-hole space,  $\int_K\equiv T\sum_n\int\frac{\dd[2]{k}}{(2\pi)^2}$, $\hat G^{b_i}_K=[\im\omega_n-\hat H^{b_i}_{\vec k}]^{-1}$, and $\hat H^{b_i}=(\epsilon^{b_i}_{\vec k}-\mu)\hat\sigma_3+\Delta_{b_i}\hat\sigma_1$. The eRS response itself can be computed from
the general formalism outlined in Refs.~\cite{maiti_etal:2016:ProbingPairingInteractionMultiple,maiti_etal:2017:ConservationLawsVertexCorrections}. The Raman spectral function in the irrep $R$ is computed as Im$[\chi_R(\Omega)]$, where (see SM; also, we suppress the Raman shift $\Omega$ for brevity)
\begin{align}\label{Eq:chi_B1g}
  \chi_R= -\hat\gamma\left[\hat\Pi_{33}+\hat\Pi_{32}\{2[\hat V_{R}^{\rm pp}]^{-1}-\hat\Pi_{22}\}^{-1}\hat\Pi_{23}\right]\hat\gamma^T,
\end{align}
where $\hat\gamma\equiv(\gamma^R_{b_1},\gamma^R_{b_2},\ldots)$ is a vector of the Raman vertices in the $R^{\rm th}$ irrep and $\hat\Pi_{mn}\equiv\diag[\Pi^{b_1}_{mn}(\Omega),\Pi^{b_2}_{mn}(\Omega),\ldots]$ are various correlations \footnote{For $\theta_{\vec k}$-dependent gaps, the $\Pi$'s have to be dressed with appropriate form factors, vide Ref.~\cite{maiti_hirschfeld:2015:CollectiveModesSuperconductorsCompeting}}. Let us apply this formula to a 2-band system with interactions
\begin{align}\label{eq:pairing_int}
  \hat V^{pp}_{\Aog}=\begin{bmatrix}
                       U^{\Aog}_{b_1}&V^{\Aog}\\
                       V^{\Aog}&U^{\Aog}_{b_2}
                     \end{bmatrix} \;\text{and}\;
  \hat V^{pp}_{\Bog}=\begin{bmatrix}
                       U^{\Bog}_{b_1}&V^{\Bog}\\
                       V^{\Bog}&U^{\Bog}_{b_2}
                     \end{bmatrix}.
\end{align}
First, consider the case with identical band parameters such that $U^R_{b_1}=U^R_{b_2}$, $V^R_{b_1}=V^R_{b_2}$ and $|\gamma^R_{b_1}|=|\gamma^R_{b_1}|$. This would lead to a ground state with $\Delta_{b_2}=-\sgn(V^{\Aog})\Delta_{b_1}$. The response in the $R^{\rm th}$ irrep then evaluates to (see SM) $\chi_R =-(\Pi_{33}/2)\times$
\begin{align}\label{eq:ideal_resp}
  \left[(\gamma^R_{b_1}+s\gamma^R_{b_2})^2\frac{m^R_{+s}}{m^R_{+s}+\chi} + (\gamma^R_{b_1}-s\gamma^R_{b_2})^2\frac{m^R_{-s}}{m^R_{-s}+\chi}\right],
\end{align}
where $\Pi_{33}\equiv\Pi^{b_1}_{33}=\Pi^{b_2}_{33}$, $\chi\equiv-2\nu_F[\Omega/(2\Delta)]^2I(\Omega)$, $\nu_F$ is the density of states at the Fermi level, $\Delta=|\Delta_{b_i}|$, and $m^R_{\pm s}\equiv 2/(U^{\Aog}-|V^{\Aog}|)-2/(U^{R}\mp |V^{R}|)$. Here we have introduced $s\equiv \sgn(V^{R}V^{\Aog})=-\sgn(V^{R}\Delta_{b_1}\Delta_{b_2})$. The function $I(\Omega)\sim1$ for $\Omega$ away from the $2|\Delta|$ edge threshold. The two terms above have poles that represent collective modes with mass $\propto\sqrt{m^R_{\pm s}}$ \footnote{See SM Sec IV C for interpretation of different values of $m^R$.}. To find the fluctuating degrees of freedom corresponding to these modes, we leverage the fact that the Raman vertex poles are the same as the ones of the linear response kernel for fluctuations of the amplitude, phase and density of SCs \cite{sarkar_maiti:2023:ElectronicRamanResponseSuperconductor}. Whilst the eigenvalues of the kernel give the collective mode frequencies, its eigenfunctions give the strengths of the fluctuations of various sectors. Through a straightforward analysis (see SM) we find that (i) the $m^R_{\pm s}$ mode has the form factor $\variation\varphi^R_{b_1}\pm s \variation\varphi^R_{b_2}$ (which means that the $m^R_+$ mode is the in-phase mode and $m^R_-$ mode is the out-of-phase one), and (ii) $m^R_{+s}$ is the lower energy mode.

It is clear from Eq.~(\ref{eq:ideal_resp}) that the spectral weight of the modes is controlled by the probe vertices (here Raman), which can lead to complete suppression of certain modes. For instance, since $m^R_{\pm}$ couple to $\gamma^R_{b_1}\pm\gamma^R_{b_2}$, when $|\gamma^R_{b_1}|=|\gamma^R_{b_2}|$, it dictates the spectral weight to be associated either with the $m^R_+$ or the $m^R_-$ mode. This is the manifestation of the selection rule in eRS. Observe that if $V^{R}=0$ then $m^R_+=m^R_-$ and the poles now appear as independent contributions from the bands with weights $\propto (\gamma^R_{b_1})^2$ and  $(\gamma^R_{b_2})^2$ and this is how much of the data on eRS have been modelled in the past.

\paragraph*{Consequences for $R=\Aog$.} Here $s=1$ and, hence, $m^{\Aog}_+$ is the low energy mode with the form $\variation\varphi_{b_1}+\variation\varphi_{b_2}$ (although $m^{\Aog}_+=0$). This is nothing but the in-phase Bogoliubov--Anderson--Goldstone (BAG) mode of the SC \footnote{This mode gets lifted to the plasmon due to the Coulomb interaction.}. On the other hand, for the higher energy mode $m^{\Aog}_-$ the form factor is $\variation\varphi_{b_1}-\variation\varphi_{b_2}$ and $m^{\Aog}_-=4|V^{\Aog}|/[(U^{\Aog})^2-(V^{\Aog})^2]$, which precisely corresponds to the system's out-of-phase Leggett mode \cite{leggett:1966:NumberPhaseFluctuationsTwoBandSuperconductors}. The selection rule then asserts that the contribution of the BAG mode ($m^{\Aog}_+=0$) couples to $\gamma^{\Aog}_{b_1}+\gamma^{\Aog}_{b_2}$. However, it would never have a spectral weight as it is massless. The Leggett mode ($m^{\Aog}_-\neq0$) couples to $\gamma^{\Aog}_{b_1}-\gamma^{\Aog}_{b_2}$ and will have more prominent spectral weight in a system with bands of the electron-hole (eh) type, where \(\gamma_{b_1}\gamma_{b_2}<0\), than electron-electron (ee) type \cite{klein:2010:TheoryRamanScatteringLeggett,sauer_blumberg:2010:ScreeningRamanResponseMultiband,cea_benfatto:2016:SignatureLeggettModeGroupA,maiti_etal:2017:ConservationLawsVertexCorrections}.
\begin{figure}[t]
  \includegraphics[width=1.0\linewidth]{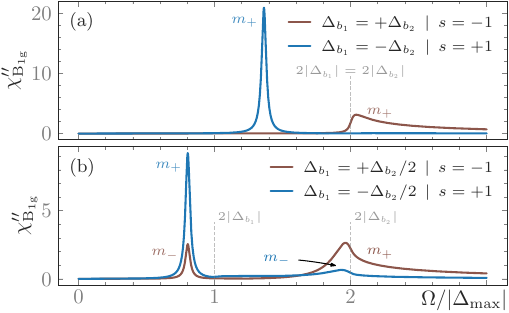}
  \phantomsubfloat{\label{fig:phasesensitive_a}}\phantomsubfloat{\label{fig:phasesensitive_b}}\vspace{-2\baselineskip}
  \caption{Phase-sensitive \Bog{} response for SCs with in- and out-of-phase gaps  with \(|V^{\Bog}|=0.9|U^{\Aog}|\) and equal Raman vertices. In (a), the equal-gaps case, the lower-energy in-gap mode is the $m_{+s} \equiv m_{+s}^{\Bog}$ mode (with $s=\pm1$). Since \( \gamma^{\Bog}_{b_1}=\gamma^{\Bog}_{b_2}\), only the $m_+$ mode couples to the eRS probe and is undamped only for $s=1$.
    (b) Different gaps leading to cross-leakage of spectral weights and hence comparable collective mode and $2\Delta$ features (for $s=-1$).
    \label{fig:phasesensitive}
  }
\end{figure}

\paragraph*{Consequences for $R=\Bog$.} In this case, the collective modes are the BaSh modes and the parameter $s=\sgn(V^{\Bog}V^{\Aog})$ is no longer fixed. The lower energy mode $m^{\Bog}_{+s}$ with form factor $\variation\varphi^{\Bog}_{b_1}+s\variation\varphi^{\Bog}_{b_1}$ will be in- or out-of-phase depending on the sign of the \Bog{} interaction as well as the sign of the \Aog{} one, which is, in turn, nothing but the sign of $-\Delta_{b_1}\Delta_{b_2}$. Thus, the \Bog{} response is inherently sensitive to sign changes of the order parameter. To understand the consequence of the selectivity let us assume, for specificity, that $\gamma^{\Bog}_{b_1}=\gamma^{\Bog}_{b_2}$. Then note the following two points: (i) the low energy $m^{\Bog}_{+s}$ mode would only be picked up by the probe for $s=1$, which corresponds to $\Delta_{b_1}\Delta_{b_2}V^{d}<0$ and not otherwise; (ii) for the case of strong interband-driven competition, $|V^{\Bog}|\gg|U^{\Bog}|$, we have $m^{\Bog}_{\pm s}=2/(U^{\Aog}-|V^{\Aog}|)\pm2/|V^{\Bog}|$. Since the \Aog{} state is the ground state, we need $|U^{\Aog}-|V^{\Aog}||>|V^{\Bog}|$, which means that $\sgn(m^{\Bog}_{\pm s})=\pm1$. This implies that a lower energy in-gap mode ($m^{\Bog}>0$) always exists while the higher energy one is damped ($m^{\Bog}<0$). However, which one will be in-phase would depend on $s$ with the eRS changing characteristically as shown in Fig.~\ref{fig:phasesensitive_a}.

\paragraph*{The selection rule beyond identical bands.} In reality no two bands are identical unless they are related by symmetry. The general form of the response is $
-\chi_R=(\gamma^R_{b_1})^2X+(\gamma^R_{b_2})^2Y+2\gamma^R_{b_1}\gamma^R_{b_2}Z$ $=(\gamma^R_{b_1}+\gamma^R_{b_2})^2(X+Y+2Z)/4$ $+(\gamma^R_{b_1}-\gamma^R_{b_2})^2(X+Y-2Z)/4$ $+[(\gamma^R_{b_1})^2-(\gamma^R_{b_2})^2](X-Y)/2$.
The particular forms of $X,Y$ and \(Z\) are (in the SM, but) not relevant except for that each of these terms contains both poles, and that $X\rightarrow Y$ under $b_1\leftrightarrow b_2$ with only $Z\propto V^{R}$. At first glance, we observe that the selectivity is lost and the response is now a mix of contributions from both modes. However, the selectivity in the \Aog{} channel remains exact as the self-consistency relations always enforce $X=Y$ and $X\pm Z$ is such that one of the poles cancels out. This couples $m^{\Aog}_\pm$ to $\gamma^{\Aog}_{b_1}\pm\gamma^{\Aog}_{b_2}$, respectively. For the \Bog{} channel the interactions are independent and the above is no longer true. However, (i) in most systems $|\gamma^R_{b_1}|\approx |\gamma^R_{b_2}|$ even if $\Delta_{b_1}\neq\Delta_{b_2}$, making the $(\gamma^R_{b_1})^2-(\gamma^R_{b_2})^2$ contribution negligible; (ii) for a large region of the parameter space of $\{U^{\Aog}_{b_1},U^{\Aog}_{b_2},V^{\Aog},U^{\Bog}_{b_1},U^{\Bog}_{b_2},V^{\Bog}\}$ the $m^{\Bog}_\pm$ modes are such that they are still largely coupled to $\gamma^{\Bog}_{b_1}\pm\gamma^{\Bog}_{b_2}$, respectively, with some ``cross-leakage''. These two points result in the selection rule being approximately valid and we refer to it as being ``fragile''. This is shown in Fig.~\ref{fig:phasesensitive_b}, in which both \(m^{\Bog}_\pm\) modes are present in the response. In fact, if we consider the case of interband-driven \Bog{} interactions where we get one in-gap mode and a damped mode near $2|\Delta|$, we see that the selection rule would have only selected either the in-gap mode or the damped mode to show up in eRS. It is precisely the ``fragility'' that allows the in-gap mode to leak into the $2|\Delta|$-dominated spectrum, giving comparable weights [see $m^{\Bog}_\pm$ for $s=-1$ in Fig.~\ref{fig:phasesensitive_b}]. This naturally explains why in materials with dominant interband interactions we can expect significant spectral weight at the largest gap even in the presence of an in-gap mode. This was an unresolved issue in Refs.~ \cite{kretzschmar_etal:2013:RamanScatteringDetectionNearlyDegenerate,bohm_etal:2014:BalancingActEvidenceStrong,wu_etal:2017:SuperconductivityElectronicFluctuationsCeBa,bohm_etal:2018:MicroscopicOriginCooperPairing}.

\begin{figure}[b]
  \includegraphics[width=1.0\linewidth]{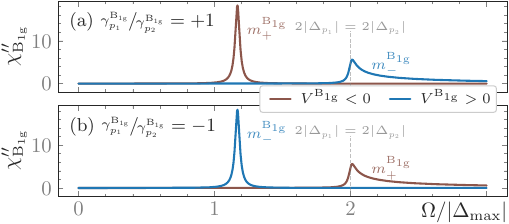}
  \phantomsubfloat{\label{fig:OffGamma_a}}\phantomsubfloat{\label{fig:OffGamma_b}}\vspace{-2\baselineskip}
  \caption{\Bog{} spectra for a 2-pocket, 1-band system with equal interactions. The ground state is given by $U^{\Aog}=V^{\Aog}<0$ and the response is different for different signs of $V^{\Bog}$. Switching the sign of $\gamma^{\Bog}_{p_1}/\gamma^{\Bog}_{p_2}$ [panels (a) and (b)] switches the in-phase/out-of-phase characteristic and vertex association of the low-energy mode.
    \label{fig:OffGamma}
  }
\end{figure}
\begin{figure*}
  \includegraphics[width=1.0\linewidth]{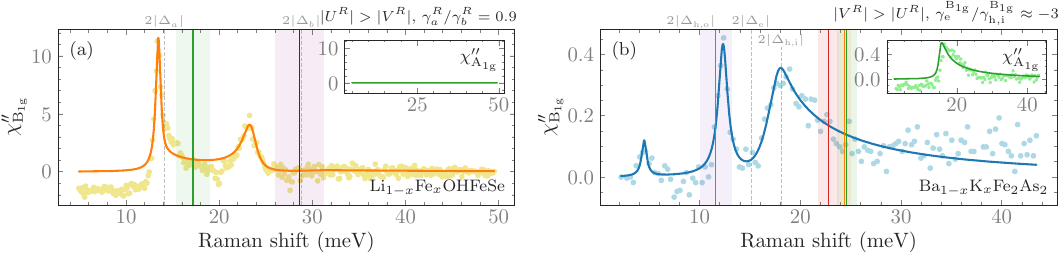}
  \phantomsubfloat{\label{fig:Exper_a}}\phantomsubfloat{\label{fig:Exper_b}}\vspace{-2\baselineskip}
  \caption{Our theoretical model fitted to (a) \LiFeOHFeSeX{} (\(x\sim 0.18)\) and (b) \BaKFeAsX{} (\(x \sim 0.48)\).
    The \Bog{} response is plotted over the experimental data reported in Refs.~\cite{he_etal:2020:RamanStudyCooperPairing} and \cite{bohm_etal:2018:MicroscopicOriginCooperPairing}, respectively, and the insets show the associated \Aog{} response. The dashed vertical lines denote the $\Delta_{b_i}$ needed to get the response, while the coloured lines (and their spread) indicates the $\Delta$-measurements (and their uncertainties) from Refs.~\cite{du_etal:2016:ScrutinizingDoubleSuperconductingGaps} and \cite{nakayama_etal:2009:SuperconductingGapSymmetryCeBa} for (a) and (b), respectively.
    \label{fig:Exper}
  }
\end{figure*}
\paragraph*{Consequences for off-$\Gamma$-point Fermi pockets.} Multiple Fermi pockets on a single band emerge from special orbital characters which constrain them to be related by symmetry. Consider the example of pockets ($p_1$ and $p_2$) being at the $\group{X}$ and $\group{Y}$ points (like in \ce{Fe}-based SCs without hybridization), where they could be viewed as two identical bands (with symmetry constraints imposed on $\hat V^{\rm pp}$ by the lattice's point group). Since this is a 1-band system, the only possible \Aog{} ground state has $\Delta_{p_1}=\Delta_{p_2}$, which could result, e.g. from $V^{\Aog}=U^{\Aog}<0$. There are no modes in \Aog{}, as this is a 1-band system, but interesting scenarios arise in the \Bog{} channel. First, note that symmetry constraints we will require $\gamma^{\Bog}_{b_1}=\pm\gamma^{\Bog}_{b_2}$. Consider first the $+$ case, where only the $\gamma^{\Bog}_{b_1}+\gamma^{\Bog}_{b_2}$ contribution is present, selecting the $m^{\Bog}_+$ mode. Thus, if $V^{\Bog}<0$, \(s=1\) and the $m^{\Bog}_+$ mode will be the lower-energy one and visible. However, if $V^{\Bog}>0$, \(s=-1\) and the $m^{\Bog}_{-}$ mode would be the low-energy one, but \textit{not be visible}. The visible higher energy $m^{\Bog}_{+}$ mode in this case has a mass $1/U^{\Aog}-2/(U^{\Bog}-|V^{\Bog}|)$. Note that for repulsive \Bog{} interactions with $U^{\Bog}\approx V^{\Bog}$, $m^{\Bog}_\pm\rightarrow\infty$, which leads to a spectral feature $\sim1/\sqrt{\Omega-2|\Delta|}$. Thus, we see that we can get characteristically different responses depending on $s=\sgn(V^{\Aog}V^{\Bog})$, even if a BaSh mode exists. This feature is not possible without multiple pockets and is demonstrated in Fig.~\ref{fig:OffGamma}. Finally, if $\gamma^{\Bog}_{b_1}=-\gamma^{\Bog}_{b_2}$, we would simply switch the mode that couples to the probe in accordance with the selection rule, vide Fig.~\ref{fig:OffGamma_b}.

\paragraph*{Explanation of experimental data.} While the appearance of novel features in the \Aog{} channel in cuprates \cite{martinho_etal:2004:OriginCeA_1g,letacon_etal:2006:InterplayCeA_1g,montiel_etal:2016:EtaCollectiveModeCe} and in the \Bog{} channel pnictides \cite{bohm_etal:2014:BalancingActEvidenceStrong,bohm_etal:2017:SuperconductivityFluctuationsCeBa1p,bohm_etal:2018:MicroscopicOriginCooperPairing} already gave important clues about the nature of SC in those materials, the line shapes and spectral weights themselves were not clearly understood. For example, in the \Bog{} response of \BaKFeAsX{}, the existing interpretation \cite{scalapino_devereaux:2009:CollectiveWaveExcitonModes,maiti_etal:2016:ProbingPairingInteractionMultiple} does not explain huge the spectral weight near $2|\Delta|$ without assuming the bands to be decoupled, which is unrealistic in interband-driven mechanisms. Similarly, the shape of the \Bog{} peaks as well as lack of \Aog{} response in \LiFeOHFeSeX{} has only been speculatively analysed using a 1-band model \cite{he_etal:2020:RamanStudyCooperPairing,scalapino_devereaux:2009:CollectiveWaveExcitonModes}.

Since we are equipped with a multiband theory \footnote{It should be noted that the formulae presented do not account for interaction contributions from the particle-hole channel, which is known to also induce collective modes \cite{khodas_etal:2014:CollectiveModesMultibandSuperconductors,maiti_etal:2017:ConservationLawsVertexCorrections}.}, we can explore these issues explicitly. In Fig.~\ref{fig:Exper_a} we show the result of fitting our model (see SM for parameters) to the \Bog{} data for \LiFeOHFeSeX{}. The fit suggests that $|U^R|\gtrsim|V^R|$ which is not unexpected for 2-pocket models \cite{du_etal:2018:SignReversalOrderParameter}.
Further, we now know that mostly the $\gamma^{\Bog}_{b_1}+\gamma^{\Bog}_{b_2}$ vertex will contribute, but since the gaps are different across the hybridized pockets, the leakage leads to coupling of both modes to this vertex. Since this is a system with only (hybridized) electron pockets at the $\group{X}$ and $ \group{Y}$ points, it leads to $\gamma^{\Aog}_{p_1}\approx\gamma^{\Aog}_{p_2}$. As the \Aog{} response couples via $\gamma^{\Aog}_{p_1}-\gamma^{\Aog}_{p_2}$ which is $\approx0$ in this system, our model explains the nearly null result in the \Aog{} channel in Ref.~\cite{he_etal:2020:RamanStudyCooperPairing} [inset of Fig.~\ref{fig:Exper_a}].

In Fig.~\ref{fig:Exper_b} we show the fit to the \Bog{} data in \BaKFeAsX{}. The parameters (see SM) suggest that $|V^R|>|U^R|$ which is expected for this system, as they are driven by interband interactions. Not only the found parameters are consistent with indications from previous microscopic calculations for the pairing \cite{graser_etal:2009:NeardegeneracySeveralPairingChannels,thomale_etal:2009:FunctionalRenormalizationgroupStudyDoping} but our ``uncontrived'' model correctly captures the spectral weights of the BaSh modes and the bumps near the $2|\Delta|$ regions. Even if we did not have the insight from microscopic calculations, we already know from this work that the presence of a strong $2|\Delta|$ feature is a signature of dominant repulsive interband interactions, which is true for this material. Without any additional fitting, we also get a finite response in the \Aog{} channel with a feature near the largest gap and no Leggett modes, which is exactly what is seen in experiments [inset of Fig.~\ref{fig:Exper_b}].

\paragraph*{Conclusion.}
Although we focused on eRS due to data availability, it should be clear to the reader that the main contribution of this work is the identification of the appropriate quantum degrees of freedom that couple to the spectroscopic probe and accounting for many-body effects that renormalize such coupling.
For eRS, we demonstrated that, in addition to the symmetry-imposed selection rules, there are also ``fragile'' many-body physics-induced ones that couple the coherent fluctuations of various degrees of freedom to the probe in a characteristic manner. We used this to simultaneously explain the spectral weights for various features in \LiFeOHFeSeX{} and \BaKFeAsX{} in both \Aog{} and \Bog{} irreps. This analysis can be readily extended to other systems that break time-reversal symmetries (vide Ref.~\cite{sarkar_maiti:2023:ElectronicRamanResponseSuperconductor}).
The new knowledge gained about the selection rules will also be relevant to emerging techniques such as current-assisted Raman spectroscopy \cite{puviani_etal:2020:CurrentassistedRamanActivationHiggsa} and optical absorption \cite{moor_etal:2017:AmplitudeHiggsModeAdmittance,nakamura_etal:2019:InfraredActivationHiggsMode}, and third-harmonic generation \cite{tsuji_aoki:2015:TheoryAndersonPseudospinResonance,cea_etal:2016:NonlinearOpticalEffectsThirdharmonic}, which all involve coupling to the sectors outlined earlier.

\paragraph*{Acknowledgements.} We thank S. Sarkar for useful conversations and also extending the work to systems with broken time-reversal symmetry. We also thank R.~Hackl for his permission to use the data shown in Fig.~\ref{fig:Exper}. This work (S.M.) was funded by the Natural Sciences and Engineering Research Council of Canada (NSERC) Grant No. RGPIN-2019-05486 and (I.B.-L.) partially via the NSERC's QSciTech CREATE programme.

\clearpage\newpage\onecolumngrid

\section*{Supplementary material for ``Many-body physics-induced selection rules: application to Raman spectroscopy''}
\setcounter{equation}{0}\renewcommand{\theequation}{S\arabic{equation}}%
\setcounter{figure}{0}\renewcommand{\thefigure}{S\arabic{figure}}%
\setcounter{table}{0}\renewcommand{\thetable}{S\arabic{table}}%

\subsection{Self-consistency equations}\label{Sec:SelfCon}
As in the main text (MT) \cite{sup_Note1}, let us consider the normal state to be comprised of concentric bands with energy dispersion $\epsilon_{\vec k}^a$ ($a$ is a band index) around the $\Gamma$ point and with chemical potential $\mu$. To address superconductivity (SC), we will focus on the singlet channel and work in the usual BCS formalism. This means that we will not account for the redundant spin degeneracy explicitly. Moreover, we will assume that the leading instability is in the \Aog{} channel and the competition is in the \Bog{} channel of a 2D square lattice. The orthogonality of the two channels allows for
a decoupled pairing interaction which can be modelled as $\hat V^{\rm pp}=\hat V^{\rm pp}_{\Aog}f^{\Aog}_{\theta_{\vec k}}+\hat V^{\rm pp}_{\Bog}f^{\Bog}_{\theta_{\vec k}}$, with form factors $f^{\Aog}_{\theta_{\vec k}}=1$ and $f^{\Bog}_{\theta_{\vec k}}=\sqrt2\cos(2\theta_{\vec k})$ and where $\theta_{\vec k}$ is the angle along the Fermi surface measured relative to the $\Gamma\group{X}$ axis. These form factors have the property that $\int\frac{\dd{\theta}}{2\pi}(f^R_{\theta_{\vec k}})^2=1$ for $R\in\{\Aog,\Bog\}$. The matrix structure of $\hat V^{\rm pp}$ represents all the intra- and interband interaction matrix elements in the particle-particle (pp) channel. For a band $a$ we introduce the order parameter $\Delta_a(\theta_{\vec k})=\Delta^{\Aog}_a f^{\Aog}_{\theta_{\vec k}}+\Delta^{\Bog}_a f^{\Bog}_{\theta_{\vec k}}$. The assumption of the leading instability being in the \Aog{} channel means the problem is defined within the interaction parameter space region where $\Delta_a^{\Bog}=0$ and $\Delta_{a}^{\Aog}=\Delta_a$ become the solution to the self-consistency equation (at temperature $T=0$). That is,
\begin{align}\label{eq:selfcon}
  \Delta_a&=-\sum_b[\hat{V}^{\rm pp}_{\Aog}]_{ab}\Delta_b\nu_F^bL_b,
            \quad
            L_b\equiv\ln\frac{2\Lambda}{|\Delta_b|},
\end{align}
where $\nu_F^b$ is the density of states (DOS) of band $b$ at the Fermi energy, $\Lambda$ is some cut-off in the Cooper problem that is defined by the pairing mechanism. We can invert the above self-consistency equation to obtain $L_a$ in terms of the order parameters, eliminating any explicit cut-off dependence,
\begin{align}\label{eq:Ls}
  \nu_F^aL_a&=-\frac{1}{\Delta_a}\sum_b[\hat V^{\rm pp}_{\Aog}]^{-1}_{ab}\Delta_b\equiv - (V^{\rm eff}_{a,\Aog})^{-1}.
\end{align}
Here we have introduced the band-wise effective pairing interaction $V^{\rm eff}_{a,\Aog}$ from band \(a\) that enters the renormalization of the Raman response in any irreducible representation (irrep).

\subsection{Correlation functions}\label{Sec:corr}
In the superconducting state there are certain correlation functions of interest [their definitions are provided in \cref{Eq:Pi33Pi32}]:
\begin{alignat}{3}\label{eq:CorrFunc}
  \text{density-density:} \quad& \Pi^a_{33}(\Omega)&&= -2\nu_F^aI_a(\Omega), \nn\\
  \text{density-phase:} \quad& \Pi^a_{32}(\Omega)&&= -2\nu_F^a\left(\frac{\im\Omega}{2\Delta_a}\right)I_a(\Omega)=-\Pi^a_{23}(\Omega) \quad\text{and}\nn\\
  \text{phase-phase:} \quad& \Pi^a_{22}(\Omega)&&=-2\nu_F^aL_a+\chi^a(\Omega),~~\chi^a(\Omega)\equiv-2\nu_F^a\left(\frac{\Omega}{2\Delta_a}\right)^2I_a(\Omega),\nn\\
  \text{where} \quad& I_a(\Omega)&&\equiv\frac{\sin^{-1}[\Omega/(2\Delta_a)]}{[\Omega/(2|\Delta_a|)]\sqrt{1-[\Omega/(2\Delta_a)]^2}} \quad\text{and}
\end{alignat}
$L_a$ is given by \cref{eq:Ls}. These correlation functions have the property that
\begin{equation}\label{eq:identity}
  \Pi^a_{33}\chi^a+(\Pi^a_{32})^2=\Pi^a_{33}\chi^a-\Pi^a_{32}\Pi^a_{23}=0.
\end{equation}

\subsection{Raman response function}\label{Sec:response}
We start from the general formula for the Raman response from a multiband system in the $R^{\rm th}$ irrep as given in Ref.~\cite{sup_maiti_etal:2017:ConservationLawsVertexCorrections}, which can be written as
\begin{align}\label{eq:RamanResponseMB0}
  -\chi_{R}&=\sum_{a}\sum_{i\in\{2,3\}}\gamma^{R*}_a\Pi^a_{3i}\Gamma^{R,a}_{i},\nn
\end{align}
where $\gamma^R_a$ is the effective-mass vertex which for a band with dispersion $\epsilon^a_{\vec k}$ is identified as follows. The $\vec k$-dependent vertex is $\gamma^R_a(\vec k)=G^R_{\alpha\beta}\partial_{k_\alpha}\partial_{k_\beta}\epsilon^a_{\vec k}$ and $\gamma^R_a=\int\frac{\dd{\theta_{\vec k}}}{2\pi}\gamma^R_a(\vec k)f^R_{\vec k}$, where $G^R_{\alpha\beta}$ are matrices that combine  $\partial_{k_\alpha}$ and $\partial_{k_\beta}$ in a way consistent with the symmetry of the $R^{\rm th}$ irrep \cite{sup_ovander:1960:FormRamanTensor}. For example, $G^{\Aog}=\diag(1,1)$, $G^{\Bog}=\diag(1,-1)$ et cetera. The index $i\in\{2,3\}$ refers to the phase and density sectors, and $\Gamma^{R,a}_i$ is the interaction-corrected Raman vertex for band $a$ which has components in both the phase and density sectors that is calculated from the vertex equation
\begin{align}
  \sum_{b,j}\left[\delta_{ab}\delta_{3i}+\frac12V^{\rm pp}_{ab}P^{b}_{ij}\right]\Gamma^{R,b}_{j}=\gamma^{R}_a\delta_{3i}, \quad\text{where}\quad P^{b}_{ij}=\begin{pmatrix}
                                                                                                                                                              \Pi^{b}_{22}&\Pi^{b}_{23}\\
                                                                                                                                                              0&0
                                                                                                                                                            \end{pmatrix}.
\end{align}
In the above formulæ, for $\chi_{R}$ we have assumed that we will only be interested in the $\vec q\to0$ response, which does not require accounting for the Coulomb interaction \cite{sup_cea_benfatto:2016:SignatureLeggettModeGroupA,sup_maiti_etal:2017:ConservationLawsVertexCorrections}. Carrying out the operations in the $\{2,3\}$-space explicitly, we arrive at
\begin{align}\label{eq:RamanResponseMB1}
  -\chi_{R}&=\sum_{a,b}\gamma^{R*}_a\left\{\Pi^a_{33}\delta_{ab}+\Pi^a_{32}\left[2[\hat V^{\rm pp}_{R}]^{-1}-\hat\Pi_{22}\right]^{-1}_{ab}\Pi^b_{23}\right\}\gamma^R_b,
\end{align}
where $\hat \Pi_{22}=\diag(\Pi_{22}^{b_1},\Pi_{22}^{b_2},\ldots)$. We then use Eq.~(\ref{eq:identity}) to replace $\Pi_{33}^a$ in terms of $\Pi^a_{32}\Pi^a_{23}$ and obtain
\begin{align}\label{eq:RamanResponseMB}
  -\chi_{R}&=\sum_{ab}\gamma^R_a\Pi^a_{32}[\hat\chi^{-1}+\hat{\mathcal P}^{-1}]_{ab}\Pi^b_{23}\gamma^R_b,\nn\\
  \text{where}\quad\hat\chi&\equiv\diag(\chi^{b_1},\chi^{b_2},\ldots),\nn\\
  \hat{\mathcal P} &= 2[\hat V^{\rm pp}_{R}]^{-1}-\hat\Pi_{22} = 2\left\{[\hat V^{\rm pp}_{R}]^{-1}-[\hat V^{\rm eff}_{\Aog}]^{-1}\right\}-\hat\chi\quad\text{and}\nn\\
  \hat V^{\rm eff}_{\Aog}&=\diag\left(V^{\rm eff}_{b_1,\Aog}, V^{\rm eff}_{b_2,\Aog},\ldots\right).
\end{align}

All of the correlation functions that enter the response function have a finite imaginary part for $\Omega>2|\Delta_a|$. The response, thus, has finite spectral weight at those frequencies. For $\Omega<2|\Delta_a|$ the spectral weight is zero unless there are resonances due to collective modes, which show up as $\delta$ function-like features in this region. In fact, the entire Raman response is best interpreted as a sum of responses from collective modes. These modes could be undamped if they are realised for $\Omega<2|\Delta_{\rm min}|$ (usually as sharp $\delta$ function-like features) or damped if realised for $\Omega>2|\Delta_{\rm min}|$ (with broad features), where $|\Delta_{\rm min}|$ is the smallest gap in a multiband system.

Observe that $(\hat V^{\rm eff}_{\Aog})^{-1}=(V^{\rm pp}_{\Aog})^{-1}$ for the 1-band case. Then a pole may arise from the zero of $2[(V^{\rm pp}_{R})^{-1}-(V^{\rm pp}_{\Aog})^{-1}]-\chi=-(m+\chi)$, where $m=2/V^{\rm pp}_{\Aog}-2/V^{\rm pp}_{R}$. We show further in \cref{Sec:Asymptote} that for small frequencies $\chi\sim -\Omega^2$.
Thus, for the 1-band response, the denominator has the form $m-\Omega^2$. For this reason, we refer to $m$ as the mass term that is responsible for the finite frequency of the modes. This form yields the well-known BaSh mode of 1-band systems \cite{sup_bardasis_schrieffer:1961:ExcitonsPlasmonsSuperconductors,sup_maiti_etal:2016:ProbingPairingInteractionMultiple,sup_maiti_etal:2017:ConservationLawsVertexCorrections} for $R=\Bog$. In general, for a multiband case, the pole condition is the existence of zeros of the $\det(\hat{\mathcal{P}})$, which can always be factored as products of $(m_a+\chi_a)$ for appropriately defined $m_a$'s. One can also see that for the multiband case where interband \Bog{} interactions are absent, the response is simply a sum of contributions of poles from $2[(V^{\rm pp}_{a,\Bog})^{-1}-(V^{\rm eff}_{a,\Aog})^{-1}]-\chi_a$. This is the same as in the original BaSh result, but with an effective leading-channel interaction from each band. These results change in multiband systems.

\subsection{Application to 2-band model}\label{Sec:2Band}
To apply the above general theory to a 2-band system (with band labels \(b_{1}\) and \(b_2\)), we start by modelling the interactions as (we change notations from the MT for brevity)
\begin{align}\label{eq:interactions}
  \hat V^{\rm pp}_{\Aog}=\begin{pmatrix}U_{b_1}^s&V^s\\V^s&U_{b_2}^s\end{pmatrix}\quad\text{and}\quad
  \hat V^{\rm pp}_{\Bog}=\begin{pmatrix}U_{b_1}&V\\V&U_{b_2}\end{pmatrix},
\end{align}
which leads to
\begin{align}\label{eq:2BEx}
  \nu_F^{b_1}L_{b_1}&=-\frac{1}{U_{b_1}^sU_{b_2}^s-(V^s)^2}\left(U^s_{b_2}-V^s\frac{\Delta_{b_2}}{\Delta_{b_1}}\right)\quad\text{and}\nn\\
  \nu_F^{b_2}L_{b_2}&=-\frac{1}{U_{b_1}^sU_{b_2}^s-(V^s)^2}\left(U^s_{b_1}-V^s\frac{\Delta_{b_1}}{\Delta_{b_2}}\right),
\end{align}
with the effective leading-channel interaction being
\begin{align}\label{eq:eff_interactions}
  \hat V^{\rm eff}_{\Aog}=\begin{pmatrix}
                            \dfrac{U_{b_1}^sU_{b_2}^s-(V^s)^2}{U^s_{b_2}-V^s\Delta_{b_2}/\Delta_{b_1}}&0\\
                            0&\dfrac{U_{b_1}^sU_{b_2}^s-(V^s)^2}{U^s_{b_1}-V^s\Delta_{b_1}/\Delta_{b_2}}
                          \end{pmatrix}.
\end{align}
The expression for $\hat{\mathcal P}^{-1}$ then becomes, with \(d \equiv \det[\hat V^{\rm pp}_{\Bog}]=U_{b_1}U_{b_2}-V^2\),
\begin{align}\label{eq:Pinv}
  \hat{\mathcal P}^{-1}&=\dfrac{1}{\mathcal D}
                         \begin{pmatrix}
                           U_{b_1}-\dfrac{d}{2}\left(\dfrac{2}{V^{\rm eff}_{b_2,\Aog}}+\chi_2\right)&V\\
                           V&U_{b_2}-\dfrac{d}{2}\left(\dfrac{2}{V^{\rm eff}_{b_1,\Aog}}+\chi_1\right)
                         \end{pmatrix},\nn\\
  \text{where}\quad
  \mathcal D &\equiv\frac{U_{b_1}U_{b_2}}{2}\left[\left(\frac2{U_{b_1}}-\frac{2}{V^{\rm eff}_{b_1,\Aog}}-\chi_{b_1}\right)\left(\frac2{U_{b_2}}-\frac{2}{V^{\rm eff}_{b_2,\Aog}}-\chi_{b_2}\right)-\frac{V^2}{U_{b_1}U_{b_2}}\left(\frac{2}{V^{\rm eff}_{b_1,\Aog}}+\chi_{b_1}\right)\left(\frac{2}{V^{\rm eff}_{b_2,\Aog}}+\chi_{b_2}\right)\right].
\end{align}
This implies that in Eq.~(\ref{eq:RamanResponseMB}) we get
\begin{align}\label{eq:Pinvchiinv}
  \hat\chi^{-1}+\hat{\mathcal P}^{-1}&= \frac{1}{\mathcal D}\begin{pmatrix}
                                                              U_{b_1}-\dfrac{d}{2}\left(\dfrac{2}{V^{\rm eff}_{b_2,\Aog}}+\chi_{b_2}\right)+\dfrac{\mathcal D}{\chi_{b_1}}&V\\
                                                              V&U_{b_2}-\dfrac{d}{2}\left(\dfrac{2}{V^{\rm eff}_{b_1,\Aog}}+\chi_{b_1}\right)+\dfrac{\mathcal D}{\chi_{b_2}}
                                                            \end{pmatrix}.
\end{align}
Due to the structure of $\hat{\mathcal P}^{-1}$, we observe that in the \Bog{} response, $\chi_{\Bog}$, it is the presence of $V$ that couples the response to the combination $\gamma^{\Bog}_{b_1}\gamma^{\Bog}_{b_2}/(\Delta_{b_1}\Delta_{b_2})$. That is, the interband interaction in the competing channel makes the response sensitive to the phase of the ground-state order parameter $\Delta$, to the attractive or repulsive nature of the interband interaction, and to the electron or hole nature of the effective-mass vertices $\gamma^{\Bog}_a$.

\subsubsection{A hidden symmetry}
It is worth noting the presence of a hidden symmetry in the expression for the Raman response $\chi_R$. It is clear from Eqs.~(\ref{eq:RamanResponseMB}), (\ref{eq:Pinv}) and (\ref{eq:Pinvchiinv}) that it is the presence of the off-diagonal structure of $\hat V^{\rm pp}_R$ (here $V$) that is responsible for mixing the responses from different bands. In fact, these off-diagonal terms between two bands \(a\) and \(b\) generates a term $\gamma^R_a\gamma^R_b\Pi^a_{23}\Pi^b_{32}\times [\hat{\mathcal P}]^{-1}_{ab}$. Therefore, since $\Pi^a_{23}\propto1/\Delta_a$, vide Eq.~(\ref{eq:CorrFunc}), we have that
\begin{align}\label{eq:HS1}
  \sgn(\text{interband \(ab\) term})=\sgn(\gamma^R_a\gamma^R_b\Delta_a\Delta_b[\hat{\mathcal P}^{-1}]_{ab}).
\end{align}
Then, the self-consistency condition in the ground state ensures that $[(V^{\rm pp}_{\Aog})^{-1}]_{ab}\Delta_b/\Delta_a$ is of a fixed sign (\(-1\)) and, hence, invariant with respect to sign flips. Furthermore, we observe that $[\hat{\mathcal P}^{-1}]_{ab}=V^{\rm pp}_{R,ab}\mathcal{F}=V\mathcal{F}$, where $\mathcal{F}$ is a function containing terms that are insensitive to $\Delta_a\rightarrow-\Delta_a$ or $V^{\rm pp}_{R,ab}\rightarrow-V^{\rm pp}_{R,ab}$. Thus, the response function is comprised of a term whose sign behaves like $\sgn(\gamma^R_{a}\gamma^R_{b}\Delta_{a}\Delta_{b}V^{\rm pp}_{R,ab})$, whereas the other terms are insensitive to the sign flips. This term can also be written as $\sgn(\gamma_{a}\gamma_{b}V^{\rm pp}_{ab,\Aog}V^{\rm pp}_{ab,\Bog})$ due to the self-consistency condition. These terms allow for certain invariances in the response with respect to pairwise sign changes of the terms involved. For example, the response is invariant under the simultaneous action of the band $b$ switching from electron-like to hole-like (sign reversal of the effective-mass vertex $\gamma_b$) and the gap $\Delta_b$ switching sign. We may refer to this invariance as a \emph{hidden symmetry} in the Raman response of superconductors. Noting that $V^{\rm pp}_{\Aog, ab}\rightarrow -V^{\rm pp}_{\Aog,ab}$ implies $\Delta_a\Delta_b\rightarrow-\Delta_a\Delta_b$, and that $\mathcal{P}^{-1}\propto V^{\rm pp}_{ab,R}=V^R$, we can reduce Eq.~(\ref{eq:HS1}) to $\sgn(\gamma^R_{a}\gamma^R_{b}V^RV^s)$. Other interesting invariances could be possible in multiband systems with 3 or more bands, but exploring that is left for future studies.

Moreover, observe that we only need $V^{\rm eff}_{a,\Aog}$ to get the \Bog{} response, the precise value of interband interaction in the ground state not being relevant to the discussion. However, we need to keep it as it controls the phase of the ground state. The interesting cases that arise are due to different values of the interactions in the \Bog{} channel, as we shall discuss below.

\subsubsection{Spectral weights}
Next, to understand and interpret the response and the spectral weight for each mode, first note the following algebraic identity, where \(\gamma_{a} \equiv \gamma^{R}\):
\begin{align}\label{eq:algebI}
  \gamma_{b_1}^2X+\gamma_{b_2}^2Y+2\gamma_{b_1}\gamma_{b_2}Z=(\gamma_{b_1}+\gamma_{b_2})^2\left(\frac{X+Y+2Z}{4}\right)+(\gamma_{b_1}-\gamma_{b_2})^2\left(\frac{X+Y-2Z}{4}\right)+(\gamma_{b_1}^2-\gamma_{b_2}^2)\left(\frac{X-Y}2\right).
\end{align}
As can be verified with an explicit calculation starting from Eq.~(\ref{eq:RamanResponseMB}), the response $-\chi_{\Bog}$ is precisely in the above form with
\begin{align}\label{eq:XYZ}
  \mathcal{D}X&=\left[U_{b_1}-\frac d2\left(\frac{2}{V^{\rm eff}_{b_2,\Aog}}+\chi_{b_2}\right)+\frac{\mathcal D}{\chi_{b_1}}\right]\Pi^{b_1}_{33}\chi_{b_1},\nn\\
  \mathcal{D}Y&=\left[U_{b_2}-\frac d2\left(\frac{2}{V^{\rm eff}_{b_1,\Aog}}+\chi_{b_1}\right)+\frac{\mathcal D}{\chi_{b_2}}\right]\Pi^{b_2}_{33}\chi_{b_2}\quad\text{and}\nn\\
  \mathcal{D}Z&=V\dfrac{\Pi^{b_1}_{32}\Pi^{b_2}_{32}+\Pi^{b_2}_{32}\Pi^{b_1}_{32}}{2}=V\Pi^{b_1}_{32}\Pi^{b_2}_{32}.
\end{align}
Note that the result in the $R^{\rm th}$ irrep is got by $\{U_{b_1},U_{b_2},V\}\rightarrow\{U^R_{b_1},U^R_{b_2},V^R\}$. Since we can get from $X$ to $Y$ by $b_1\leftrightarrow b_2$, we see that the whole expression is invariant under $b_1\leftrightarrow b_2$, as it should be in the physical case. Furthermore, $Z$ is the only term sensitive to $\sgn(V\Delta_{b_1}\Delta_{b_2})$ and it couples to $\gamma^{R}_{b_1}\gamma^{R}_{b_2}$ as expected from the hidden symmetry discussion. However, this explicit expression allows us to infer more about
the effects of changes in \(\sgn(V)\): it switches the association of $X+Y\pm2Z$ with $\gamma^{R}_{b_1}\pm\gamma^{R}_{b_2}$, i.e. switches the contribution from $\gamma^{R}_{b_1}+\gamma^{R}_{b_2}$ and $\gamma^{R}_{b_1}-\gamma^{R}_{b_2}$ terms. Let us understand the implication of this result by exploring some special cases. In \cref{eq:ideal_resp}, the presented form corresponds to setting the parameters for the two bands to be identical, which results in $X=Y$ and $\chi_a=\chi$. In the MT we present a form where the parameter $s=\sgn(V^sV)$ has been drawn out from the $\mathcal D$ and migrated to the $\gamma^R$'s and the $m^R$'s. We discuss this explicitly in Sec.~\ref{Sec:New Rule}.

\subsubsection{Case with no interband competition: $V=0$}
In this case, it is easy to see that $\hat{\mathcal P}^{-1}=-\diag[1/(m_{b_1}+\chi_{b_1}),1/(m_{b_2}+\chi_{b_2})]$, where $m_a\equiv 2/V^{\rm eff}_{a,\Aog}-2/U_a$.
This leads to
\begin{align}\label{eq:repsonse1}
  -\chi_{\Bog}=\sum_a\gamma_a^2\Pi^a_{33}\frac{m_a}{m_a+\chi_a},
\end{align}
where we have used the relation $\Pi^a_{33}\chi_a=\Pi^a_{32}\Pi^a_{23}$ and which is simply the sum of the contributions from the (independent) bands. Since $V=0$ there is no coupling of the response between the bands. Thus, the response is not sensitive to $\Delta_a\rightarrow-\Delta_a$ or $\gamma_a\rightarrow-\gamma_a$. This is the limit in which most multi-band responses have been modelled in past fits to experimental studies and that is why it was thought that the response would be blind to sign changes in $\gamma^R_a$ and $\Delta_a$. As we see from the MT and the subsequent section here in this supplementary text, doing so completely misses important interband interaction induced effects. Continuing with $V=0$, we have that (and these are well known):
\begin{itemize}
  \item If $1/m_a\rightarrow+\infty$, that is $U_a\gtrsim V^{\rm eff}_{a,\Aog}$ (or $|U_a|\lesssim |V^{\rm eff}_{a,\Aog}|$ since these interactions have to be negative in order to have competition), then we approach a \Bog-instability.
  \item If $1/m_a\rightarrow 0^{+}$, that is $U_a\rightarrow 0^{-}$ then we have a mode at $\Omega\lesssim2|\Delta_a|$. That is, attractive \Bog-interaction leads to long-lived collective modes.
  \item If $1/m_a\rightarrow 0^{-}$, that is $U_a\rightarrow 0^{+}$ then the mode at $\Omega\gtrsim2|\Delta_a|$ is damped by the continuum of excitations. That is, repulsive interaction leads to damped collective behaviour.
\end{itemize}
The $\Omega$-dependent properties of \cref{eq:repsonse1} which affects the line shape will discussed in Sec.~\ref{Sec:Asymptote} (below).

\subsubsection{Effect of interband competition: $V\neq 0$}
The presence of a finite $V$ turns the response sensitive to the phase of the $\Delta_a$, $V$ and $\gamma_a$ due to the hidden symmetry property. However, before exploring the response, let us understand the nature of the collective modes the poles represent. For the case where the response from individual bands were decoupled, the collective modes simply corresponded to independent phase fluctuations of each band. Now, in the presence of the interband interaction $V$ (in the competing channel), the phase fluctuations of the two bands are coupled leading to the formation of in-phase and out-of-phase fluctuation components. This is analogous to the Goldstone- and the Leggett-type fluctuations in the leading \Aog{} channel. The amplitude and phase fluctuations can be quantified as
\begin{align*}
  \updelta\Delta_a=\updelta[|\Delta_a|e^{\im\varphi_a}]=\underbrace{e^{\im\varphi_a}\updelta|\Delta_a|}_{\rm amplitude}+\im\underbrace{|\Delta_a|e^{\im\varphi_a}\updelta\varphi_a}_{\rm phase}.
\end{align*}
To proceed from here, we follow the linear response treatment in Refs.~\cite{sup_maiti_hirschfeld:2015:CollectiveModesSuperconductorsCompeting,sup_sarkar_maiti:2023:ElectronicRamanResponseSuperconductor}, where it is outlined how the interaction-renormalised Raman vertex is the same as the linear response kernel involving the amplitude and phase fluctuations. The linear response kernel is written by expanding the dynamical self-consistency equation to linear order in the amplitude and phase fluctuations. In time-reversal symmetric superconductors, the phase sector decouples from the amplitude sector. The eigenvalue problem for the two-band case of the linear response kernel then gives us
\begin{align}\label{eq:phases}
  \Delta_{b_1}\updelta\varphi_{b_1}=\frac{\dfrac V2\Pi^{b_2}_{22}}{1-\dfrac{U_{b_1}}2\Pi^{b_1}_{22}}\Delta_{b_2}\updelta\varphi_{b_2} \quad\implies\quad \updelta\varphi_{b_1}=\frac{\dfrac V2\dfrac{\Delta_{b_2}}{\Delta_{b_1}}\Pi^{b_2}_{22}}{1-\dfrac{U_{b_1}}2\Pi^{b_1}_{22}}\updelta\varphi_{b_2}.
\end{align}
If we focus on the sign of the terms and use the relation $\sgn(\Delta_{b_2}/\Delta_{b_1})=-\sgn(V^s)$, along with the knowledge that within the gap ($\Omega<2|\Delta|$) we have $1-U^{R}_{b_1}\Pi^{b_1}_{22}/2>0$, we obtain
\begin{align}
  \label{eq:phases2}\sgn(\updelta\varphi_{b_1}\updelta\varphi_{b_2})=\sgn(VV^s).
\end{align}
Noting that $\sgn(V_s)=-\sgn(\Delta_{b_1}\Delta_{b_2})$ we can introduce $s\equiv \sgn(V^{R}V^{s})=- \sgn(V^{R}\Delta_{b_1}\Delta_{b_2})=\sgn(\updelta\varphi_{b_1}\updelta\varphi_{b_2})$ and create the form factor $\updelta\varphi_{b_1}+ s\updelta\varphi_{b_2}$ for the fluctuations which correspond to the modes with masses $m^R_{s}$. The in-phase mode has $\updelta\varphi_{b_1}\updelta\varphi_{b_2}>0$ whereas an out-of-phase mode has $\updelta\varphi_{b_1}\updelta\varphi_{b_2}<0$.

\subsection{A new selection rule}\label{Sec:New Rule}
Starting from the general formulae in Eqs.~(\ref{eq:algebI}) and (\ref{eq:XYZ}), let us consider the ideal case with identical Fermi surfaces where $|\Delta_{b_1}|=|\Delta_{b_2}|$, $U_{b_1}=U_{b_2}=U$, $V^{\rm eff}_{b_1,\Aog}=V^{\rm eff}_{b_2,\Aog}=V_{\Aog}$. This leads to $\chi_{b_1}=\chi_{b_2}=\chi$. We also have $|\gamma_{b_1}|=|\gamma_{b_2}|$. To account for the phase of the ground state, let us also introduce a factor $f\in\{-1,1\}$ such that $\Delta_{b_2}=f\Delta_{b_1}=f\Delta$, where $f=-\sgn(V^s)$, as required by the self-consistency equations. This leads to $V_{\Aog}=U^s+fV^s$ and $\Pi^{b_1}_{32}=f\Pi^{b_2}_{32}$. In this limit
\begin{align*}
  \mathcal{D}=\frac{d}{2}\left(\frac2{U+V}-c\right)\left(\frac2{U-V}-c\right),\quad\text{where}\quad c\equiv\frac{2}{V_{\Aog}}+\chi,
\end{align*}
which, in turn, leads to
\begin{align}\label{eq:repponse_selectivity}
  -\chi_{\Bog}&=(\gamma_{b_1}^2+\gamma_{b_2}^2)\left[\chi\left(U-\frac d2c\right)+\mathcal D\right]\frac{\Pi_{33}}{\mathcal D}+2\gamma_{b_1}\gamma_{b_2}\left[Vf\chi\right]\frac{\Pi_{33}}{\mathcal D}\nn\\
              &=\left[(\gamma_{b_1}+\gamma_{b_2})^2\left(\frac1{V_{\Aog}}-\frac{1}{U+fV}\right)\left(c-\frac2{U-fV}\right) + (\gamma_{b_1}-\gamma_{b_2})^2\left(\frac1{V_{\Aog}}-\frac{1}{U-fV}\right)\left(c-\frac2{U+fV}\right)\right]\frac{d\Pi_{33}}{2\mathcal D}\nn\\
              &=\left[(\gamma_{b_1}+\gamma_{b_2})^2\left(\frac{m_+}{m_++\chi}\right)+(\gamma_{b_1}-\gamma_{b_2})^2\left(\frac{m_-}{m_-+\chi}\right)\right]\frac{\Pi_{33}}2,
\end{align}
where $m_{\pm}=2/V_{\Aog}-2/(U\pm fV)$. Let us now note a couple of points: (i) The mode frequencies ($\propto m_\pm$) map to each other under $V^s\rightarrow-V^s$ or $V\rightarrow -V$ and hence the resonance location is not affected under these flips. But their weights switch between $(\gamma^R_a\pm\gamma^R_b)^2$. (ii) Since $f=-\sgn(V^s)$, we can write $U\pm fV=U\mp s |V|$, where $s=\sgn(VV^s)$. These remarks lead us to
\[m_{\pm}=\frac{2}{V_{\Aog}}-\frac{2}{U\mp s|V|}\quad\implies\quad m_{\pm s}=\frac{2}{V_{\Aog}}-\frac{2}{U\mp |V|}.\]
Introducing $m^R_{\pm s}$ also requires changing $\gamma^R_{b_1}\pm \gamma^R_{b_2}$ to $\gamma^R_{b_1}\pm s\gamma^R_{b_2}$. This is the form presented in the MT and \emph{represents the selection rule which couples the $m_\pm$ mode to the $\gamma^R_{b_1}\pm\gamma^R_{b_2}$ combinations of the probes}.

Moreover, analytically, for systems with different gaps across bands, the approximate selectivity will still be found. It is, after all, nothing but a switch of spectral weight that is also evident from Eq.~(\ref{eq:algebI}) where switching the sign of $Z$ [$\equiv\sgn(fV)$] flips the association with $\gamma^{R}_{b_1}\pm\gamma^{R}_{b_2}$, whilst maintaining a common background arising from the ${\gamma^{R}_{b_1}}^2-{\gamma^{R}_{b_2}}^2$ term.

\subsection{Asymptotic expansions and spectral weights}\label{Sec:Asymptote}
In this section we perform asymptotic analysis of our expressions. It helps us understand the building blocks from which one can understand the coupled response.

\subsubsection{1-band case}
\begin{figure}[htb]
  \centering\includegraphics{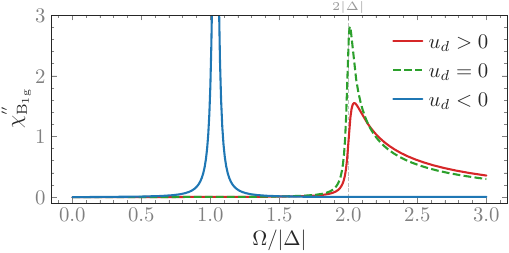}
  \caption{1-band \Bog{} response of an \swave{} SC, \(u_s=-1\), for different signs of \(|u_d|=0.9\) (solid lines) and in the absence of \Bog{} interactions (dashed line). For repulsive \(u_d\), the \(2|\Delta|\) feature softens, having smaller spectral weight when compared to the \(u_d=0\) case. For attractive \(u_d\), a BaSh mode is found in the gap and no feature is found at \(2|\Delta|\). Note that no spectral weight is \emph{transferred} (see text).}
  \label{fig:1Bandvariations}
\end{figure}

Let us investigate the possible line shapes of BaSh modes in 1-band systems. The spectrum is given by $\chi_{\Bog}=-\gamma^2\Pi_{33}m/(m+\chi)$, where $m=2/U_s-2/U$. Using the forms of $\Pi_{33}$ and $\chi$ previously given, and introducing dimensionless interactions \(u_{s}\equiv \nu_FU^s\) and \(u_{d} \equiv \nu_F U\), we get
\begin{equation}\label{eq:asymp1B}
  \chi_{\Bog}=2\nu_F\gamma^2\frac{\check m I}{\check m-\left(\dfrac{\Omega}{2\Delta}\right)^2I},~~\text{where}~\check m = \frac{1}{u_s}-\frac{1}{u_d}.
\end{equation}
The function $I$, defined in Eq.~(\ref{eq:CorrFunc}), has the form $I\approx \mathcal{O}(1)+1/\sqrt{1-\Omega/(2|\Delta|)}\equiv I_R+\im I_I$ for $\Omega\approx 2|\Delta|$. Since $I_I$ is finite only for $\Omega>2|\Delta|$, the spectrum, which is actually the $\IM(\chi_{\Bog})$, evaluates to
\begin{equation}\label{eq:AsymptotResponse1B}
  \IM(\chi_{\Bog})=2\check m^2\frac{I_I}{(I_R-\check m)^2+I_I^2}.
\end{equation}
It is clear that $\IM(\chi_{\Bog})\approx0$ for $\Omega<2|\Delta|$ with the finite weight due to the imaginary part of $\Omega+\im/\tau$. This small weight blows up to a $\delta$-function if, for some $\Omega$, we get $I_R\sim\Omega^2=\check m$. This is the condition for the presence of a BaSh mode. Since $I_R>0$, we need $\check m=1/u_{s}-1/u_{d}>0$ to have a BaSh mode. For $u_{s}$ to lead to the ground state and for $u_{d}$ to be competitive, we need $u_{s}<0$ and \(u_{d}<0\). This means that we need $|u_{d}|<|u_{s}|$ to get the BaSh mode. This is the conventional result. Note that if $u_{d}\rightarrow 0$ then $\check m\rightarrow\infty$ and $\IM(\chi_{\Bog})=I_I$ which behaves are $1/\sqrt{\Omega/(2|\Delta|)-1}$ for $\Omega\gtrsim2|\Delta|$, displaying an edge singularity. For finite $\check m$, two cases arise:
\begin{itemize}
  \item There is a BaSh mode [$\check m>0$, that is, $1/m\in(0,\infty)$]: It exists for $\Omega<2|\Delta|$. Near the BaSh mode's frequency \(\Omega_{\mathrm{BaSh}}\) we have $\IM(\chi_{\Bog})\sim \check m^2\delta(\Omega-\Omega_{\mathrm{BaSh}})$. Near and above the onset of the continuum, we get $\IM(\chi_{\Bog})\sim \check m^2\sqrt{\Omega/(2|\Delta|)-1}$. That is, the response starts from zero and develops as $\sqrt{x}$ with the prefactor~$\check m^2$.
  \item There is a damped BaSh mode ($\check m<0$, that is, $1/m<0$): The response is zero for $\Omega<2|\Delta|$. Near and above $2|\Delta|$ we once again have $\IM(\chi_{\Bog})\sim \check m^2\sqrt{\Omega/2|\Delta|-1}$.
\end{itemize}
Although the response near $2|\Delta|$ has the same functional form and prefactor, what is different in the two cases is the value of $\check m=1/u_{s}-1/u_{d}$. To have superconductivity, one needs $u_{s}<0$. Thus, $|\check m|=|1/|u_{s}|+1/u_{d}|$. For repulsive interaction in the competing channel, $u_{d}>0$, $|\check m|$ is a larger number compared to when the interaction is attractive, $u_{d}<0$, which also leads to a BaSh mode. This is why, in the 1-band case, the spectral weight ($\propto\check m^2$) near $2|\Delta|$ reduces in the presence of a BaSh mode, whereas it increases in the absence of it [vide Fig.~\ref{fig:1Bandvariations}].
This phenomenon has been colloquially referred to as the collective mode ``stealing" the spectral weight from the continuum, which is not technically correct as there is no sum rule protecting the spectral weight. And this is also the limitation of modelling systems as copies of one-band models: you either get spectral weight near $2|\Delta|$ or at the collective mode but not both.

\subsubsection{2-band case}
For all those cases where the response could be seen as a combination of decoupled poles (either for $V=0$ or for identical bands), its form still resembles $m/(m+\chi)$ and the same analysis works, with the difference being that each pole can now have its own $\Delta$. The pole with the effective attractive interaction will host a BaSh mode with reduced spectral weight near $2|\Delta|$ whilst the one with a repulsive interaction will have a larger spectral weight near $2|\Delta|$. In these cases, the only thing that is different from the 1-band case is that, in a 2-band model, we can simultaneously have attractive and repulsive channels leading to simultaneous presence of BaSh modes and spectral weights near $2|\Delta|$.

If we start from the case with only intraband interactions in the \Bog{} channel, where there are two collective modes, and introduce a weak interband component, $|V|\ll|U_a|$, then the mode frequencies repel each other \cite{sup_maiti_etal:2016:ProbingPairingInteractionMultiple}. Depending on the parameters, one of them could enter the in-between-the-gaps continuum of excitations and be damped with a rate $\propto V^2$. This is evident by solving the equation $\det(\mathcal{P})=0$ in this limit.
Furthermore, the fragile selection rule will bestow the spectral weight either entirely on the collective mode or on the $2|\Delta|$ region.

For strong interband-driven competition, $|U|\ll|V|$, we have $m_{\pm s}=2/V_{\Aog}\mp2/|V|$. Since $|V_{\Aog}|>|V|$, we get $\sgn(m_{\pm s})=\pm$. This means that we shall always have one attractive (and, hence, in the gap) and one repulsive (and, hence, damped near $2|\Delta|$) modes. Once again, the attractive or repulsive nature of this interaction will switch the spectral weights between the mode and the $2|\Delta|$ region.

Furthermore, note that, in the multiband case, the contribution for $\Omega>2|\Delta_a|$ is always of the type $\sqrt{\Omega-2|\Delta_a|}$ for each band as well. For the smallest gap, the response starts from zero and for the largest gap, the same response develops on the background of the spectral weight from the continuum of the band with the smaller gap. This contribution (from the threshold of the continuum) is almost never dominant. This was also pointed out in Ref.~\cite{sup_maiti_etal:2017:ConservationLawsVertexCorrections}. This is significantly different from the non-interacting case where the threshold has a $1/\sqrt{\Omega-2|\Delta|}$ singularity.

\subsection{Experimental data and theoretical curves parameters}

In this section, we elaborate on how we have fitted the experimental data of \LiFeOHFeSeX{} and \BaKFeAsX{} using the 2-band theoretical framework described in this work \cite{sup_Note2}. To that end we used SciPy v1.11.3's \cite{sup_scipy1_0contributors_etal:2020:SciPyFundamentalAlgorithmsScientific} non-linear least squares \texttt{curve\_fit} method (with unchanged defaults) and assumed the difference spectra \(\to 0\) in the high-energy limit. We sampled, for each SC, different regions of the 2-band, \Aog{} ground state parameter space defined by: the interaction parameters, the SC gaps, the Raman vertices projections, the lifetime effect (disorder) parameter \(\eta\) and an overall scale as to match the experimental data. \(\eta\) comes from analytical continuation: the frequencies \(\Omega\) in the MT are replaced, for the corresponding bands \(a\), by \(\Omega+\im\eta_a\). Also, an initial guess for the parameters was made with a set whose resulting response would have features that would loosely match the experiment.

Furthermore, based on experiments and known physical constraints, we placed the following constraints to our parameter space: (i) we imposed an upper bound for the gap values based on data reported by Ref.~\cite{sup_du_etal:2016:ScrutinizingDoubleSuperconductingGaps} for \LiFeOHFeSeX{} and Ref.~\cite{sup_nakayama_etal:2009:SuperconductingGapSymmetryCeBa} for \BaKFeAsX{}, and (ii) the \(\sgn(\gamma_{a}^{R})\) was fixed to be \(+1\) (\(-1\)) for electron (hole) bands.

\subsubsection{\LiFeOHFeSeX}

The dimensionless interaction matrices, vide \cref{Sec:Asymptote}, found in our modelling process are
\begin{equation*}
  \nu_{F}\hat{V}_{\Aog}^{pp} \equiv
  \begin{bmatrix}
    -0.951 & -0.419 \\
    -0.419 & -0.951 \\
  \end{bmatrix}
  \quad\quad\text{and}\quad\quad
  \nu_{F}\hat{V}_{\Bog}^{pp} \equiv
  \begin{bmatrix}
    -0.819 & -0.240 \\
    -0.240 & -0.819 \\
  \end{bmatrix},
\end{equation*}
where we assume the density of states to be equal across bands and the rows and columns are ordered following the order parameter hierarchy:
\begin{alignat*}{2}
  \Delta_{a} &= \SI{+7.06}{\milli\eV} &\quad\text{and}\\
  \Delta_{b} &= \SI{+14.4}{\milli\eV}.
\end{alignat*}

Because the pockets are related by symmetry, the Raman vertices are expected to be equal in magnitude. However, due to the nature of the hybridised electron pockets of this system, when sampling the parameter space, we allowed the Raman vertices to change, being limited to be within 10\% of each other.

The additional parameters found in our modelling process and used to produce \cref{fig:Exper_a} are compiled in \cref{tab:exp_fit_fese}. Note that the difference response for the \Aog{} channel is not displayed in the inset of \cref{fig:Exper_a} because it is trivial, except for a residual line associated with a phonon shift and present due to high spectral resolution \cite{sup_he_etal:2020:RamanStudyCooperPairing}.
\begin{table}[htbp]
  \caption{Additional parameters, not mentioned in the text, of the model used to fit the difference spectra of \LiFeOHFeSeX{} (\(x\sim0.18\)).}%
  \label{tab:exp_fit_fese}
  \begin{tabular}[t]{LS}
    \toprule
    \text{Quantity}       & {Value} \\
    \midrule
    \gamma^{\Bog}_{a}     & 0.900 \\
    \gamma^{\Bog}_{b}     & 1.00 \\
    \midrule
    \gamma^{\Aog}_{a}     & 0.900 \\
    \gamma^{\Aog}_{b}     & 1.00 \\
    \midrule
    \eta                  & 0.0225 \(|\Delta_{b}|\) \\
    \text{Overall factor} & 1.99 \\
    \botrule
  \end{tabular}
\end{table}

\subsubsection{\BaKFeAsX{}}

The dimensionless interaction matrices found in our modelling process are
\begin{equation*}
  \nu_{F}\hat{V}_{\Aog}^{pp} \equiv
  \begin{bmatrix}
    -1.00        & \phantom{+}0 & \phantom{+}0 \\
    \phantom{+}0 & -0.300       & +0.866 \\
    \phantom{+}0 & +0.866       & -0.300
  \end{bmatrix}
  \quad\quad\text{and}\quad\quad
  \nu_{F}\hat{V}_{\Bog}^{pp} \equiv
  \begin{bmatrix}
    -0.974       & \phantom{+}0 & \phantom{+}0 \\
    \phantom{+}0 & \phantom{+}0 & +1.00 \\
    \phantom{+}0 & +1.00        & \phantom{+}0
  \end{bmatrix},
\end{equation*}
where we assume the density of states to be equal across bands and the rows and columns are ordered following the order parameter hierarchy:
\begin{alignat*}{3}
  \Delta_{\mathrm{h,o}} &= \SI{+5.80}{\milli\eV} &\quad\text{(outer hole pocket)},\\
  \Delta_{\mathrm{e}}   &= \SI{-7.55}{\milli\eV} &\quad\text{(electron pocket)}\phantom{,} & \quad\text{and}\\
  \Delta_{\mathrm{h,i}} &= \SI{+9.05}{\milli\eV} &\quad\text{(inner hole pocket)}.
\end{alignat*}

When sampling the parameter space, the Raman vertex ratio \(\gamma_{\mathrm{e}}^{\Bog}/\gamma_{\mathrm{h,i}}^{\Bog}\) was limited to \(3\), as larger ratios are unrealistic. Furthermore, although the outer hole pocket is decoupled from the other bands, we modelled a residual interband coupling by increasing the lower bound of \(\eta_{\mathrm{e}} = \eta_{\mathrm{h,i}}\) to be at least 25\% larger than \(\eta_{\mathrm{h,o}}\), as their spectral features lay on the continuum of excitations of the band hosting the outer hole pocket.

\emph{No other constraints were imposed in our modelling} and the \Aog{} response was calculated using the parameters found from the \Bog{} fit, except for its Raman vertices, \(\gamma^{\Aog}\), which were independently chosen.
The additional parameters found in our modelling process and used to produce \cref{fig:Exper_b} are compiled in \cref{tab:exp_fit_feas}.

It is worth reiterating that \emph{only} the \Bog{} data was used in the fitting process. And we found a set of parameters with dominant interband interactions, consistent with microscopic calculations for the pairing mechanism in this material, which correctly captures the spectral weights in both \Bog{} \emph{and} \Aog{} channels, vide \cref{fig:Exper_b} and its inset.
\begin{table}[htbp]
  \caption{Additional parameters, not mentioned in the text, of the model used to fit the difference spectra of \BaKFeAsX{} (\(x\sim0.48\)).}%
  \label{tab:exp_fit_feas}
  \begin{tabular}[t]{LS}
    \toprule
    \text{Quantity}                         & {Value} \\
    \midrule
    \gamma^{\Bog}_{\mathrm{h,o}}            & -0.336 \\
    \gamma^{\Bog}_{\mathrm{e}}              & 1.00 \\
    \gamma^{\Bog}_{\mathrm{h,i}}            & -0.333 \\
    \midrule
    \gamma^{\Aog}_{\mathrm{h,o}}            & 1.00 \\
    \gamma^{\Aog}_{\mathrm{e}}              & 1.45 \\
    \gamma^{\Aog}_{\mathrm{h,i}}            & -1.45 \\
    \midrule
    \eta_{\mathrm{h,o}}                     & 0.0500 \(|\Delta_{\mathrm{h,o}}|\) \\
    \eta_{\mathrm{e}} = \eta_{\mathrm{h,i}} & 0.0625 \(|\Delta_{\mathrm{h,i}}|\) \\
    \text{Overall factor}                   & 0.422 \\
    \botrule
  \end{tabular}
\end{table}

\end{document}